\documentclass[prb,preprint,amsmath,amssymb]{revtex4}
\begin{document}
\author{Susan B. Rempe,$^*$ Thomas R. Mattsson,  and K. Leung}
\affiliation{Sandia National Laboratories, MS 1415, Albuquerque, NM 87185}
\title{On "the complete basis set limit" and plane-wave methods in
first-principles simulations of water}

\input epsf
 
\begin{abstract}
Water structure, measured by the height of the first peak in oxygen-oxygen
radial distributions, is convverged with respect to plane-wave basis energy
cutoffs for {\it ab initio} molecular dynamics simulatinos, confirming the
reliability of plane-wave methods.
\end{abstract}

\maketitle

Given its ubiquitous presence both in our external and internal environments,
and its active participation in numerous chemical processes, 
water is arguably the most important of all liquids.  Yet despite
extensive study and documentation water has been called the 
least understood of the known liquids
with many of its properties considered anomalous.\cite{franks}

In the thirty-five years since Franks' comprehensive assessment,\cite{franks} 
much progress has been made toward attaining a molecular-level understanding
af water, particularly with regard to water structure.\cite{head-gordon}
High accuracy scattering data collected over this time, including the most recent studies,\cite{hura} 
converge on a view that water is modestly
structured.  Accurate structural data enables quantitative assessment of
computer simulations, which potentially can play a pivotal role in advancing
our understanding not only of water and its anomalies, but also of
solvation and interfacial water behavior.\cite{head-gordon} 
We focus this study on {\it ab initio} molecular dynamics (AIMD) 
simulations of water and assess whether structural predictions are
converged with respect to a particular choice of methodology.

Most AIMD simulations of water\cite{klein,pratt,grossman,galli,kuo,vande,marzari,voth,martyna} using the BLYP\cite{blyp} and PBE\cite{pbe} functionals
report enhanced structure of liquid water relative to the most recent
X-ray scattering data.\cite{hura} Over-structured  water is predicted
even though these simulations use disparate methods including varied
pseudo-potentials, and simulation protocols.
An exception occurs when the RPBE\cite{rpbe} exchange-correlation functional 
is applied, for which the predicted water structure nearly matches
experiment.\cite{pratt}  
All of these simulations cited share one common feature: expansion of the
wavefunctions in a plane-wave basis set.

A recent AIMD study by Lee and Tuckerman\cite{tuck1} (henceforth referred to
as LT) apply an alternative basis set and report a less significant 
over-structuring of water relative to most prior simulations.  
Specifically, LT find that their simulations using the BLYP exchange-correlation functional\cite{blyp} 
yield an O-O pair correlation function $g(r)$
below 3.0 at temperature 300~K,
indicating a less significant over-structuring than earlier works.\cite{klein,pratt,grossman,galli,kuo,vande,marzari,voth,martyna} 
The authors employ a real-space method to expand the
wavefunctions, Discrete Variable Representation (DVR), and stress
the high degree of energy and force convergence achieved using
that basis set. LT also speculated that
effects of an incomplete plane wave basis set and consequent
inadequate force convergence in previous work\cite{klein,pratt,grossman,galli,kuo,vande,marzari,voth,martyna}
could in general be
responsible for overstructured water $g(r)$ published in the
literature for multiple functionals, including not just BLYP but also
PBE and PW91.\cite{tuck1}

The discrepancy between DVR and existing plane wave predictions
for BLYP water, attributed to under convergence of plane wave basis
used in the literature,\cite{tuck1} has motivated us to examine this
issue for another functional widely used for water, namely PBE.
We ask the question of what EC is required to converge
water structure for AIMD simulations using plane wave basis set expansions.
Toward this end, we present plane-wave based AIMD simulations of
water (D$_2$O), and systematically analyze convergence of 
the maximum height of the first peak in the oxygen-oxygen $g(r)$ with respect 
to the
wavefunction energy cutoff (EC). 

The simulations are made using
the code VASP.\cite{vasp} We study cells with 64 and 32 molecules at
1.0~g/cc density (assuming, for this purpose only, the proton mass)
in the NVT (Nose-Hoover thermostat) ensemble at
300~K and 375~K, starting from several different initial configurations  
with $\Gamma$-point Brillouin zone sampling.  We employ as a general model the PBE\cite{pbe}
 exchange-correlation functional with
plane-augmented waves applied to describe the interaction
between core and valence electrons, a pseudo-potential which has been used
extensively for many systems.  The Born-Oppenheimer energy convergence
criteria, time steps (0.25 or 0.5 ps) used, and total energy drifts\cite{note}
in these runs are listed in Table~1.  

Equilibration protocols used
are these:  run A started from the last configuration of a
simulation using the empirical SPC/E\cite{spce} force field (2~ps equilibration); 
B, from the end of A (0~ps); C, from the 
end of B (0~ps); D, from the end of A (6.2~ps); E, from a T=400~K
simulation (12~ps); 
F, 8~ps into E (3~ps); G, end of E (0~ps); H, 8~ps into E (5~ps).

Figure~1 illustrates equilibration of the thermostat for simulations conducted
on 32 waters at temperatures of 375~K (trajectory B) and 300~K (trajectory D).  
($v$) of each atom type ($X=$O or H, for $N=$32 oxygens or 64 protons,
using the deuterium mass for protons), scaled by Boltzmann's constant
($k_{\rm B}$) and system temperature ($T$):
\begin{equation}
T = \sum_{X=1}^{X=N}(1/2)m_X|v_X|^2/(3k_{\rm B}/2).
\end{equation}

Equilibration of both O and H atoms
to the same temperature is achieved within 3-4~ps of simulation time using a single global
Nose-Hoover thermostat.  During the 12~ps equilibration of run E to ambient temperature from a high-temperature initial
configuration (T=400~K),
the $g(r)$ peak value remained between 3.0 and 3.4 (not illustrated), in other words over-structured, until equilibrium was established.


Figure~2a shows the O-O and H-H pair correlation functions for
D$_2$O at 375~K,  with peak heights further listed in Table~1.  
At this temperature, the pair-correlation functions do not become
less overstructured when EC is increased by a factor of two beyond
the VASP recommended energy cutoff of 400 eV for our oxygen and
hydrogen PAW pseudopotentials.  (Note that PAW simulations generally
require a lower EC than norm-conserving pseudopotentials to yield
converged water $g(r)$.)

At 300~K (see Fig.~2b) the
slow dynamics of PBE water cause larger statistical errors in O-O peak height
compared to the 375~K runs. 
The width and
rise of the distribution functions are, in contrast, well converged. 
Faster dynamics in the 375~K runs result in more efficiently sampled simulations,
making simulations at 375~K  more relevant demonstrations of
peak-height convergence with EC due to this statistical effect.  Nevertheless, there is no reason to conclude
that convergence behavior of AIMD simulations with respect to EC would
be different at 375~K and at 300~K;  the difference in thermal
energy between the two temperatures is in fact negligible in comparison
to the energy fluctuations present in an NVT simulation at these
temperatures.

Our $g_{\rm OO}(r)$ computed at T=300~K (Fig.~2b) are similar to those
of Ref.~\onlinecite{pratt} and~\onlinecite{galli}, the former 
of which was performed at a slightly higher temperature of T=337~K.
Considering the statistical uncertainties, it is more
significant that at these temperatures the PBE functional
consistently predicts considerable overstructuring of water.

The case of EC=300 eV (Fig.~2b, Trajectory F in Table~1) is chosen to
demonstrate the kind of deviations caused by a too low EC.
For example, $g_{\rm OO}(r)$ appears significantly less over-structured
than converged, EC=400 or 500~eV simulations,
and its first peak is shifted to a smaller $r$ value at this EC.
At least for the PBE functional, our observed trend does not
support LT's speculation that {\it raising} EC to improve convergence
may yield {\it less} overstructured water.

A possible source of errors in AIMD simulations is the fidelity
of the pseudo-potentials.  We have verified that employing a harder
set of potentials (PAW with nominal 700 eV cutoff) does not lead to
changes in $g(r)$.  

Finally,
given the smooth and asymptotic character of convergence
with respect to EC (Fig.~1 in Ref.~\onlinecite{tuck1}),
it is highly unlikely that going well beyond EC=1000 eV
could result in any changes in structural properties. 
 
Our main conclusion is thus that it is crucial to assess convergence
of the property of interest (here $g_{\rm OO}(r)$) in any kind of AIMD
simulation.
The property needs to be converged with respect to, for example,
plane-wave energy cut-off, as well as fictitious mass, $k$-point, type and
hardness of pseudo-potentials, time-step, and system size.
Our independent and well-converged calculations of the O-O pair correlation
function for water, conducted while systematically
varying EC and enforcing identical pseudopotentials,
Brillouin zone sampling, and simulation protocol (Born-Oppenheimer
as opposed to Car-Parrinello molecular dynamics), indicate that
raising EC does not lessen overstructuring of PBE water at 1.0 g/cc density.
In contrast, convergence studies of different properties, like energy or pressure (see
for example Ref.~\onlinecite{leung} and \onlinecite{mattsson}), may or may not result in
a converged structure.

To summarize, although the results of Lee and Tuckerman are
interesting, and are sure to stimulate additional work on this
important subject, we have shown that
the overstructured water $g(r)$'s 
computed at T$~\sim$~300~K using the PBE functional published in the
literature are not artifacts due to underconverged plane-wave cutoffs.

This work was supported by Sandia's LDRD program.  Sandia is a multiprogram laboratory
operated by Sandia Corporation, a Lockheed Martin Company, for the
U.S.~Department of Energy's National Nuclear Security Administration
under contract DE-AC04-94AL8500.


\newpage

\begin{figure}
\centerline{\hbox{\epsfxsize=3.20in \epsfbox{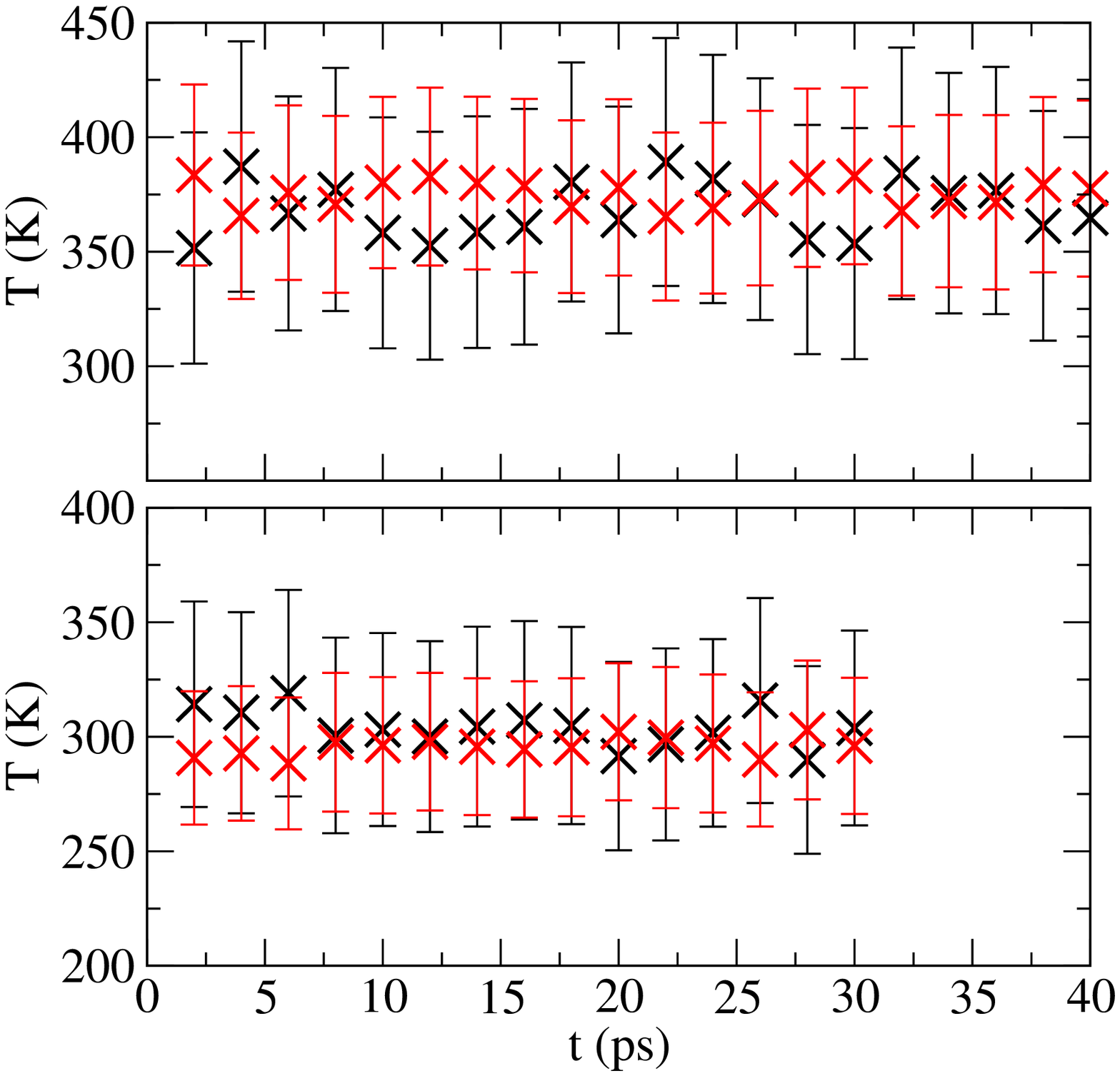}}}
\caption[]
{\label{kin} \noindent
Atomic temperatures (T) in units Kelvin averaged for 2~ps time segments over 32 oxygen (O, depicted in black) and 64 hydrogen (X, red) atoms during the
time course (t) in ps of a) Trajectory B, b) Trajectory D.  The error
bars represent one standard deviation in temperature for each atomic
species at each time step.
}
\end{figure}

\begin{figure}
\centerline{\hbox{\epsfxsize=3.20in \epsfbox{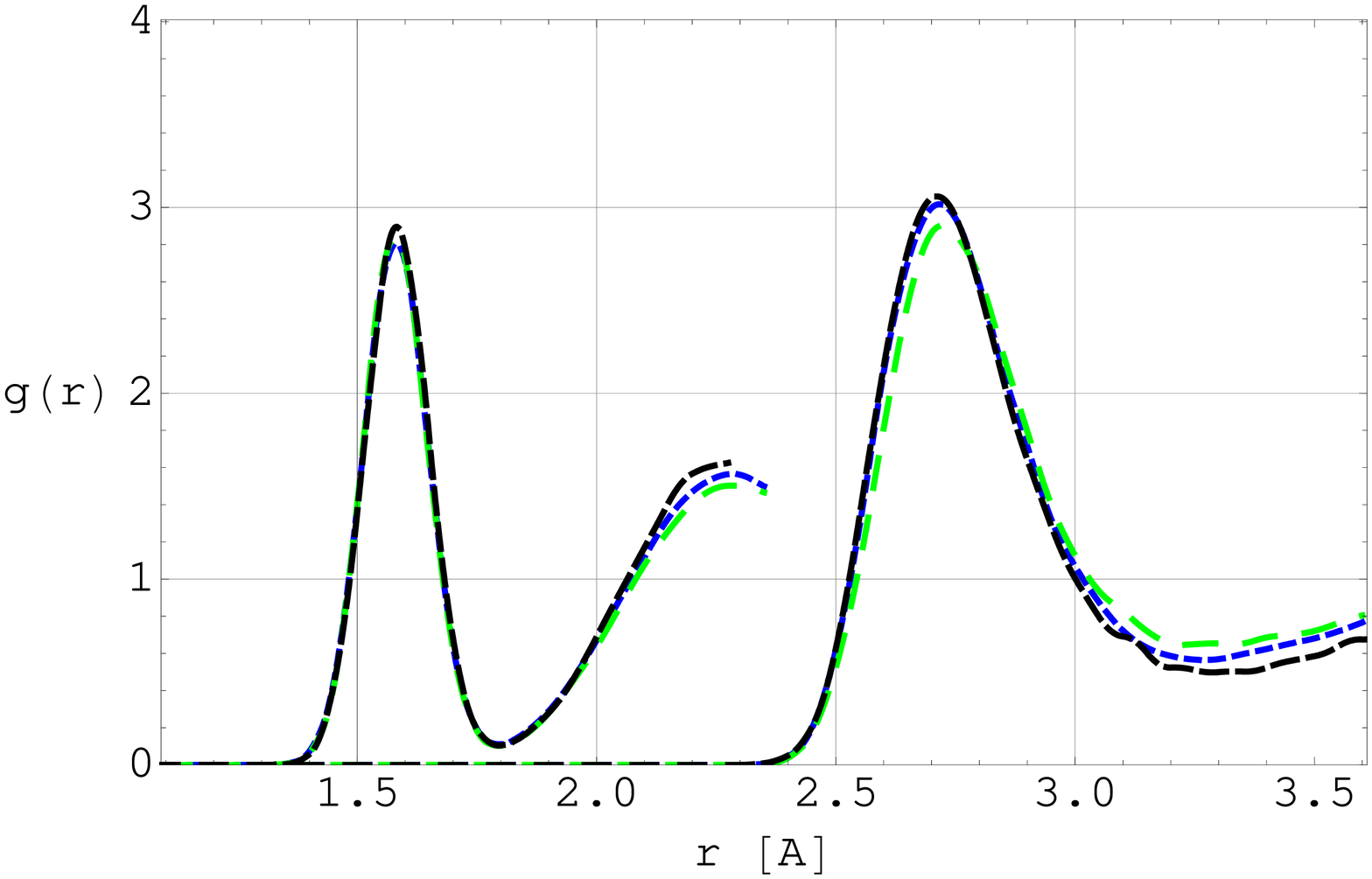}}(a)}
\centerline{\hbox{\epsfxsize=3.20in \epsfbox{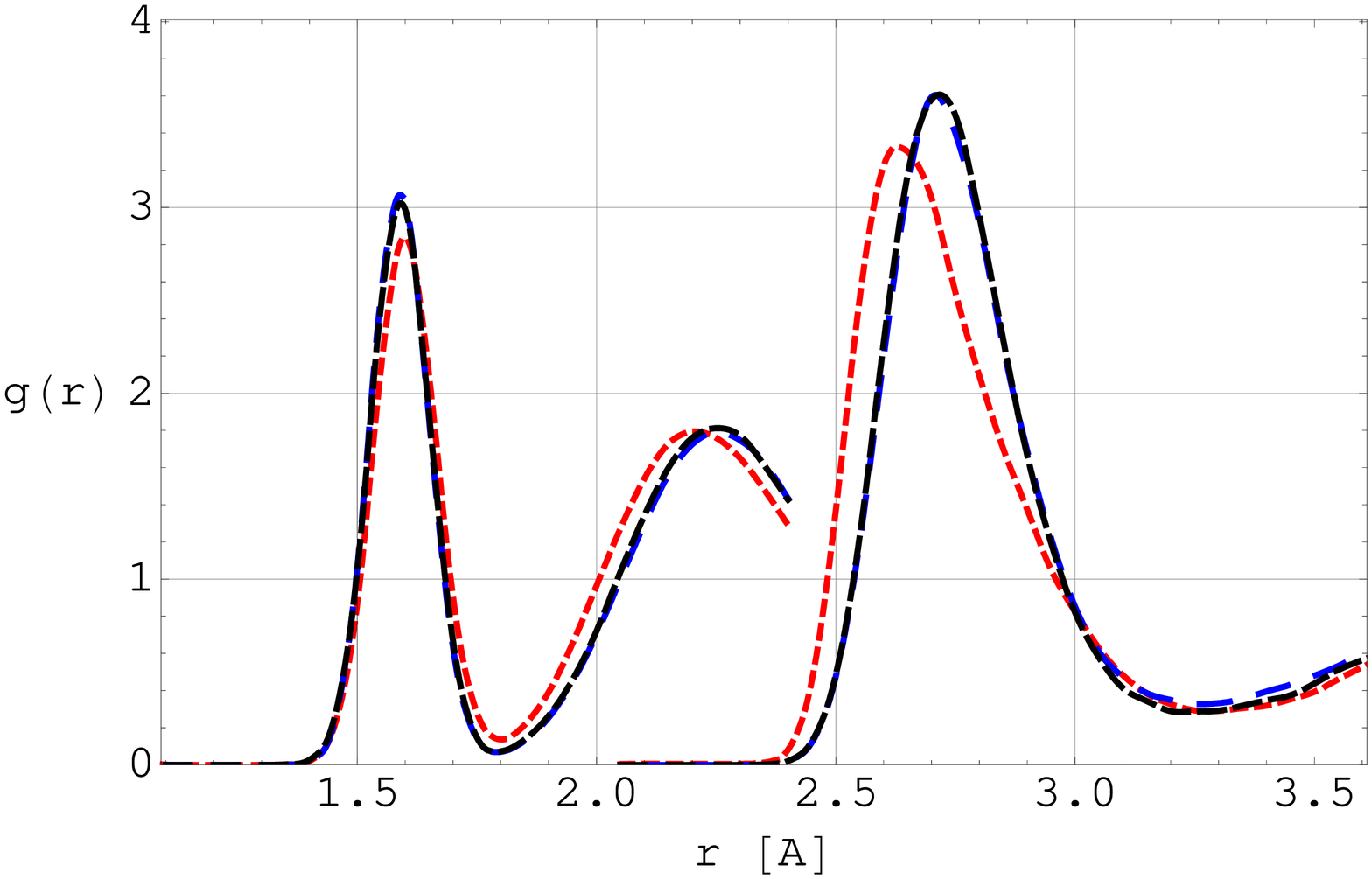}}(b)}
\caption[]
{\label{fig1} \noindent
Pair-correlation functions $g_{\rm HH}(r)$ (at shorter distances) and $g_{\rm OO}(r)$ 
(at longer distances) for D$_2$O.
(a) 32 molecule cell at 375~K: EC=400 (green), 800 (blue), and 1200~eV (black) (runs A, B, C in Table~1).
(b) 64 molecules cell at 300~K: EC=300 (red), 500 (blue), 1000 (black) (runs F, G, H in Table~1).
}
\end{figure}

\begin{table}\centering
\begin{tabular}{||l|l|r|r|r|r|r|r|c||} \hline
traj. &
N$_{\rm water}$ & EC (eV) & temp (K) & conv & $t$ (ps) & $\Delta t$ (ps)
& $g_{\rm OO}(r)_{\rm max}$ &
drift (K/ps) \\ \hline
A & 32 & 400 & 375 & 10$^{-6}$ & 24 & 0.50 & 2.92$\pm$0.07 & 1.6 \\
B & 32 & 800 & 375 & 10$^{-7}$ & 40 & 0.25 & 3.10$\pm$0.10 & 0.6 \\
C & 32 & 1200 & 375 & 10$^{-7}$ & 32 & 0.25 & 3.05$\pm$0.10 & 0.7 \\
D & 32 & 800 & 300 & 10$^{-6}$ & 29 & 0.25 & 3.75$\pm$0.10 & 1.5 \\ 
E & 64 & 500  & 300 & 10$^{-4}$ & 29 & 0.5 & 3.67$\pm$0.15 & 3.8 \\
F & 64 & 300  & 300 & 10$^{-5}$ & 12 & 0.5 & 3.40$\pm$0.15 & 2.5 \\
G & 64 & 500  & 300 & 10$^{-6}$ & 7.5 & 0.5 & 3.67$\pm$0.15 & 0.1 \\
H & 64 & 1000 & 300 & 10$^{-5}$ & 15 & 0.5 & 3.57$\pm$0.15 & 0.1 \\
\hline
\end{tabular}
\caption[]
{\label{table1} \noindent
Details of our AIMD/PBE simulations. Each trajectory ({\it traj.}) is identified
by an alphabetical label, the number of waters ($N_{water}$) in the simulation box,
the wavefunction energy cutoff in eV ({\it EC}), the thermostat temperature in Kelvin ({\it temp}),
the energy convergence tolerance between successive self-consistent iterations during
electronic structure calculations ({\it conv}),  simulation time in ps ($t$), timestep in
ps ($\Delta t$), maximum value of the first peak in the oxygen-oxygen radial distribution function
($g_{\rm OO}(r)_{\rm max}$), and the drift in temperature per ps ({\it drift}).
}
\end{table}


\begin{references}

\bibitem{franks}
 {\it Water:  A comprehensive treatise}, Franks, F., Ed.; Plenum Press:
New York, 1972.

\bibitem{head-gordon}
T. Head-Gordon and G. Hura, {\it Chem. Rev.}, 
2002, {\bf 102}, 2651.

\bibitem{hura}
G. Hura, J.M. Sorenson, R.M. Glaeser,  T. Head-Gordon , {\it J. Chem. Phys.}, 
2000, {\bf 113}, 9140.

\bibitem{klein}
B. Chen, I. Ivanov, J.M. Park, M. Parrinello, and M.L. Klein, {\it J. Phys. Chem. B}, 
2002, {\bf 106}, 12006.

\bibitem{pratt}
D. Asthagiri, L.R. Pratt, and J.D. Kress, {\it Phys. Rev. E}, 
2003, {\bf 68}, 041505.

\bibitem{grossman}
J.G. Grossman, E. Schwegler, E.W. Draeger, F. Gigy, and G. Galli, {\it J. Chem. Phys.}, 
2004, {\bf 120}, 300.

\bibitem{galli}
E. Schwegler, J.C. Grossman, F. Gygi, and G. Galli, {\it J. Chem. Phys.}, 
2004, {\bf 121}, 5400.

\bibitem{kuo}
I.W. Kuo, C.J. Mundy, M.J. McGrath {\it et al.}, {\it J. Phys. Chem. B}, 
2004, {\bf 108}, 12990.


\bibitem{vande}
J. VandeVondele, F. Mohamed, M. Krack, J. Hutter, and M. Parrinello, {\it J. Chem. Phys.}, 
2005, {\bf 122}, 014515.

\bibitem{marzari}
P.H.-L. Sit and N. Marzari, {\it J. Chem. Phys.}, 
2005, {\bf 122}, 204510.

\bibitem{voth}
S. Izvekov and G.A. Voth, {\it J. Chem. Phys.}, 
2005, {\bf 123}, 044505.

\bibitem{martyna}
Y.A. Mantz, B. Chen, and G.J. Martyna, {\it J. Phys. Chem. B}, 
2006, {\bf 110}, 3540.

\bibitem{blyp}
A.D. Becke, {\it Phys. Rev. A}, 
1988, {\bf 38}, 3098.

\bibitem{pbe}
J.P. Perdew, K. Burke, and M. Ernzerhof, {\it Phys. Rev. Lett.}, 
1996, {\bf 77}, 3865.

\bibitem{rpbe}
B.Hammer, L.B.~HansenL. and J.K.~Norskov, {\it Phys.~Rev.~B}
1999, {\bf 59}, 7413.

\bibitem{tuck1}
H.S. Lee and M. Tuckerman, {\it J. Chem. Phys.}, 2006, {\bf 125}, 154507;
see also H.S. Lee and M. Tuckerman, {\it J. Chem. Phys.}, 2007, {\bf 125},
164501 for dynamical properties, which is not the focus of our short
communication.

\bibitem{vasp}
G. Kresse and D. Jourbert, {\it  Phys. Rev. B} 1999, {\bf 59}, 1758, and references therein.

\bibitem{note}
Although the thermostat maintains the set average temperature, errors in
the forces and time integration lead to a drift in energy being absorbed
by the thermostat.  As a measure of the errors in time integration,
we convert the drift in total energy, including the energy of the
thermostat, to the equivalent in temperature, based on 1/2 $k_{\rm B}T$
energy per degree of freedom.

\bibitem{spce}
H.J. Berendsen, J.R. Grigera, and T.P. Straatsma, {\it  J. Phys. Chem.} 1987, {\bf 91}, 6269.


\bibitem{leung}
K. Leung and S.B. Rempe, {\it Phys. Chem. Chem. Phys.}, 
2006, {\bf 8}, 2153.

\bibitem{mattsson}
T.R. Mattsson and M.P. Desjarlais, {\it Phys. Rev. Lett.}, 
2006, {\bf 97}, 017801.



\end{references}
\end{document}